# Induced solitons formed by cross polarization coupling in a birefringent cavity fiber laser

**H. Zhang, D. Y. Tang*, and L. M. Zhao**

School of Electrical and Electronic Engineering,
Nanyang Technological University, Singapore 639798

**H. Y. Tam**

Department of Electrical Engineering,
Hong Kong Polytechnic University, Hong Kong, China

*: edytang@ntu.edu.sg, corresponding author.

We report on the experimental observation of induced solitons in a passively mode-locked fiber ring laser with birefringence cavity. Due to the cross coupling between the two orthogonal polarization components of the laser, it was found that if a soliton was formed along one cavity polarization axis, a weak soliton was also induced along the orthogonal polarization axis, and depending on the net cavity birefringence, the induced soliton could either have the same or different center wavelengths to that of the inducing soliton. Moreover, the induced soliton always had the same group velocity as that of the inducing soliton. They form a vector soliton in the cavity. Numerical simulations confirmed the experimental observations.

Optical solitons were first experimentally observed in single mode fibers (SMF) by Mollenauer et al. in 1980 [1]. The formation of the solitons was a result of the balanced interaction between the effects of anomalous fiber dispersion and the pulse self-phase modulation (SPM). To observe the solitons it is necessary that the intensity of a pulse be above a threshold where the nonlinear length of the pulse becomes comparable with the dispersion length. Apart from SPM, theoretical studies have also shown that cross-phase modulation (XPM) could lead to soliton formation. Soliton formation through XPM was known as the induced soliton formation. An important potential application of the effect is the light controlling light. Various cases of soliton formation in SMF caused by XPM were theoretically predicted [2-5]. Spatial soliton formation through XPM has been experimentally observed [6]. However, to the best of our knowledge, no induced temporal soliton formation experiments have been reported. In this letter, we report on the experimental observation of induced solitons in a passively mode-locked fiber laser.

Our fiber laser is schematically shown in Fig. 1. It is an erbium-doped fiber (EDF) ring laser mode-locked with a semiconductor saturable absorber mirror (SESAM). The ring cavity consists of a piece of 4.6 m EDF with group velocity dispersion (GVD) parameter 10 ps/km/nm, and 5.4 m standard SMF with GVD parameter 18 ps/km/nm. A polarization independent circulator was used in the cavity to force the unidirectional operation of the ring and simultaneously incorporate the SESAM in the cavity. Note that within one cavity round-trip the pulse propagates twice in the SMF between the circulator and the SESAM. A 10% fiber coupler was used to output the signals, and the laser was pumped by a high power Fiber Raman Laser source (BWC-FL-1480-1) of wavelength 1480 nm. The SESAM used was made based on GalnNAs quantum wells. It has a saturable absorption modulation depth of 5%, a saturation fluence of 90μJ/cm$^2$ and 10 ps relaxation time. The central absorption wavelength of the SESAM is at 1550nm. An intra cavity polarization controller was used to change the cavity linear birefringence.

As no polarizer was used in the cavity, due to the birefringence of the fibers the cavity exhibited obvious birefringence features, e. g. varying the linear cavity birefringence we could observe either the group velocity locked vector soliton (GVLVS) [7], a vector soliton whose two orthogonal polarization components have different soliton frequencies, or the polarization locked vector soliton (PLVS), characterized by that the phases of the two orthogonal polarization components of the vector soliton are locked [8]. As there is no measureable difference on the RF spectrum between the two vector soliton laser operations, in order to identify features of the vector solitons, we further explicitly investigated their polarization resolved spectra under various experimental conditions. We let the laser output first pass through a rotatable external cavity polarizer, based on the measured soliton intensity change with the orientation of the polarizer we then identify the long and short polarization ellipse axes of the vector solitons. In our measurements we found that apart from vector solitons with comparable coupled orthogonal polarization components, vector solitons with very asymmetric component intensity also exist. Fig. 2 shows for example two cases experimentally observed. Fig. 2a shows a case measured under laser operation with a relatively large cavity birefringence. In this case the spectral intensity difference between the two orthogonal polarization directions at the center soliton wavelength is more than 30 dB. The soliton nature of the strong polarization component is obvious as characterized by the existence of the Kelly sidebands. However, we emphasize that Kelly sidebands have also appeared on the weak polarization component. It shows that the weak component also is a soliton. The first order Kelly sidebands of the strong component locates at 1547.4 nm and 1569.5nm, respectively; those of the weak component are at 1546.8nm and 1568.9nm. The separations of both sets of sidebands are the same. The different locations of their Kelly sidebands of the solitons exclude the possibility that the sidebands of the weak component were produced due to an experimental artifact.

Using a commercial autocorrelator we measured the soliton pulse width. Assuming a $Sech^2$ pulse profile it is about 1 ps. If only the SPM is considered, we estimate that the peak power of the fundamental solitons in our laser is about 24W. This is well in agreement with the experimentally measured peak power of 25W of the strong component soliton. The experimentally measured peak power of the weak component soliton is only 0.5W. Obviously with the intensity of the weak pulse it is impossible to form a soliton. The weak soliton should be an induced soliton.

Through adjusting the intra cavity polarization controller the net cavity birefringence could be changed. Another situation as shown in Fig. 2b where both solitons have the same center wavelength was also obtained. Even in the case the two solitons have different Kelly sidebands. In particular, the first order Kelly sidebands of the weak soliton have slightly larger separation than that of the strong soliton, indicating that the induced soliton has a narrower pulse width than that of the strong soliton [9].

To confirm the experimental observation, we numerically simulated the operation of the laser. We used the following coupled Ginzburg-Landau equations to describe the pulse propagation in the weakly birefringent fibers:

$$\begin{cases} \frac{\partial u}{\partial z} = i\beta u - \delta\frac{\partial u}{\partial t} - \frac{ik''}{2}\frac{\partial^2 u}{\partial t^2} + \frac{ik'''}{6}\frac{\partial^3 u}{\partial t^3} + i\gamma(|u|^2 + \frac{2}{3}|v|^2)u + \frac{i\gamma}{3}v^2u^* + \frac{g}{2}u + \frac{g}{2\Omega_g^2}\frac{\partial^2 u}{\partial t^2} \\ \frac{\partial v}{\partial z} = -i\beta v + \delta\frac{\partial v}{\partial t} - \frac{ik''}{2}\frac{\partial^2 v}{\partial t^2} + \frac{ik'''}{6}\frac{\partial^3 v}{\partial t^3} + i\gamma(|v|^2 + \frac{2}{3}|u|^2)v + \frac{i\gamma}{3}u^2v^* + \frac{g}{2}v + \frac{g}{2\Omega_g^2}\frac{\partial^2 v}{\partial t^2} \end{cases} \quad (1)$$

Where, u and v are the normalized envelopes of the optical pulses along the two orthogonal polarized modes of the optical fiber. $2\beta = 2\pi\Delta n/\lambda = 2\pi/L_b$ is the wave-number difference between the two modes. $2\delta = 2\beta\lambda/2\pi c$ is the inverse group velocity difference. $k''$ is the second order dispersion coefficient, $k'''$ is the third order dispersion coefficient and $\gamma$ represents the nonlinearity of the fiber. $g$ is the saturable gain coefficient of the fiber and $\Omega_g$ is the bandwidth of the laser gain. For undoped fibers $g$=0; for erbium doped fiber, we considered its gain saturation as

$$g = G\exp[-\frac{\int(|u|^2 + |v|^2)dt}{P_{sat}}] \quad (2)$$

where $G$ is the small signal gain coefficient and $P_{sat}$ is the normalized saturation energy.

The saturable absorption of the SESAM is described by the rate equation [10]:

$$\frac{\partial l_s}{\partial t} = -\frac{l_s - l_0}{T_{rec}} - \frac{|u|^2 + |v|^2}{E_{sat}} l_s \quad (3)$$

Where $T_{rec}$ is the absorption recovery time, $l_0$ is the initial absorption of the absorber, and $E_{sat}$ is the absorber saturation energy. To make the simulation possibly close to the experimental situation, we used the following parameters: $\gamma$=3 $W^{-1}km^{-1}$, $\Omega_g$ =24nm, $P_{sat}$=100 pJ, $k''_{SMF}$= -23 $ps^2$/km, $k''_{EDF}$= -13 $ps^2$/km, $k'''$= -0.13 $ps^3$/km, $E_{sat}$=0.6 nJ, $l_0$=0.1, and $T_{rec}$ = 2 ps, Cavity length L= 10 m.

Fig. 3 shows the simulations obtained under different cavity birefringence. Fig. 3 (a) shows the case of the laser with a beat length $L_b$ = 0.1m. In this case the strong soliton can either be formed along the slow or the fast axis of the cavity. Associated with the strong soliton there is always a weak soliton induced in the orthogonal polarization direction. The induced soliton has different central wavelength from the strong soliton, which causes that the Kelly sidebands of them have different locations. However, the wavelength shifts of their sidebands to the central soliton wavelength are the same, just like the experimental observation. Fig. 3 (b) shows a case of the cavity with a beat length of $L_b$=10m. Due to the small cavity birefringence, the induced soliton has always exactly the same central wavelength as the strong soliton. Numerically we found that as $L_b$ changed from 10m to 100m, the strong soliton swapped from the slow axis to the fast axis of the cavity as a result of the polarization instability [11]. Nevertheless, in both cases the first order Kelly sidebands of the weak soliton have slightly larger separation than that of the strong soliton.

To verify that the weak soliton is induced by the strong soliton through XPM, the XPM terms were deliberately removed from the simulations. It was found that in this case the weak component kept continuously fading away, no stable soliton pulse could be formed. We note that apart from the Kelly sidebands, in Fig. 2 and Fig. 3 there are some other discrete sharp spectral peaks. We have numerically identified their formation as caused by the four-wave-mixing (FWM) between the two polarization-components [12]. By removing the FWM terms from Equation (1), these spectral sidebands completely disappeared from the numerically calculated soliton spectra. Like the experimental observations, the FWM sidebands are more pronounced on the spectra of the induced solitons due to their weak intensity. Especially, the first order of the FWM sidebands appeared between the first order Kelly sidebands and the central soliton wavelength, and their exact positions varied with the cavity linear birefringence.

The numerical simulation has well reproduced the experimental observations. Moreover, it shows that independent of the cavity birefringence, induced solitons could always be formed in the laser. In particular, in the case of large cavity birefringence the induced soliton forms a GVLVS with the inducing soliton, where due to the different frequencies of the soliton components, their Kelly sidebands have different locations. In the case of weak cavity birefringence, a PLVS is formed. As the phases of the two soliton components are locked and the weak component is induced by the strong one, the induced soliton has narrower pulse width. Consequently, its first order Kelly sidebands have larger spectral separation. Finally, we note that the PLVS observed in our experiment is a special case of those reported in [8]. A subtle difference between them is that the two polarization components of the PLVSs reported in [8] have comparable strength. Therefore there is mutual XPM between the two soliton components. There is no such mutual XPM in the current case.

In conclusion, formation of induced temporal solitons has been experimentally observed in a passively mode-locked fiber laser with birefringence cavity. It was found that the induced solitons were formed by the XPM between the two orthogonal polarization components of the birefringence laser, and the induced solitons could either have the same or different soliton frequency to the inducing soliton. As the induced solitons always have the same group velocity as that of the inducing soliton, they form vector solitons in the laser. To our knowledge, this is the first experimental observation of temporal induced solitons.

## Figure caption:

Fig. 1: Schematic of the experimental setup. EDF: Erbium doped fiber. WDM: wavelength division multiplexer.

Fig. 2: Polarization resolved vector soliton spectra experimentally measured. (a): Obtained under large cavity birefringence. (b) Obtained under relatively weaker cavity birefringence. The corresponding RF spectra after a polarizer are given in the figure 2(a).

Fig. 3: Polarization resolved vector soliton spectra numerically calculated. (a): $L_b$= 0.1m. (d) $L_b$= 10m. Pump strength $G_0$=80.

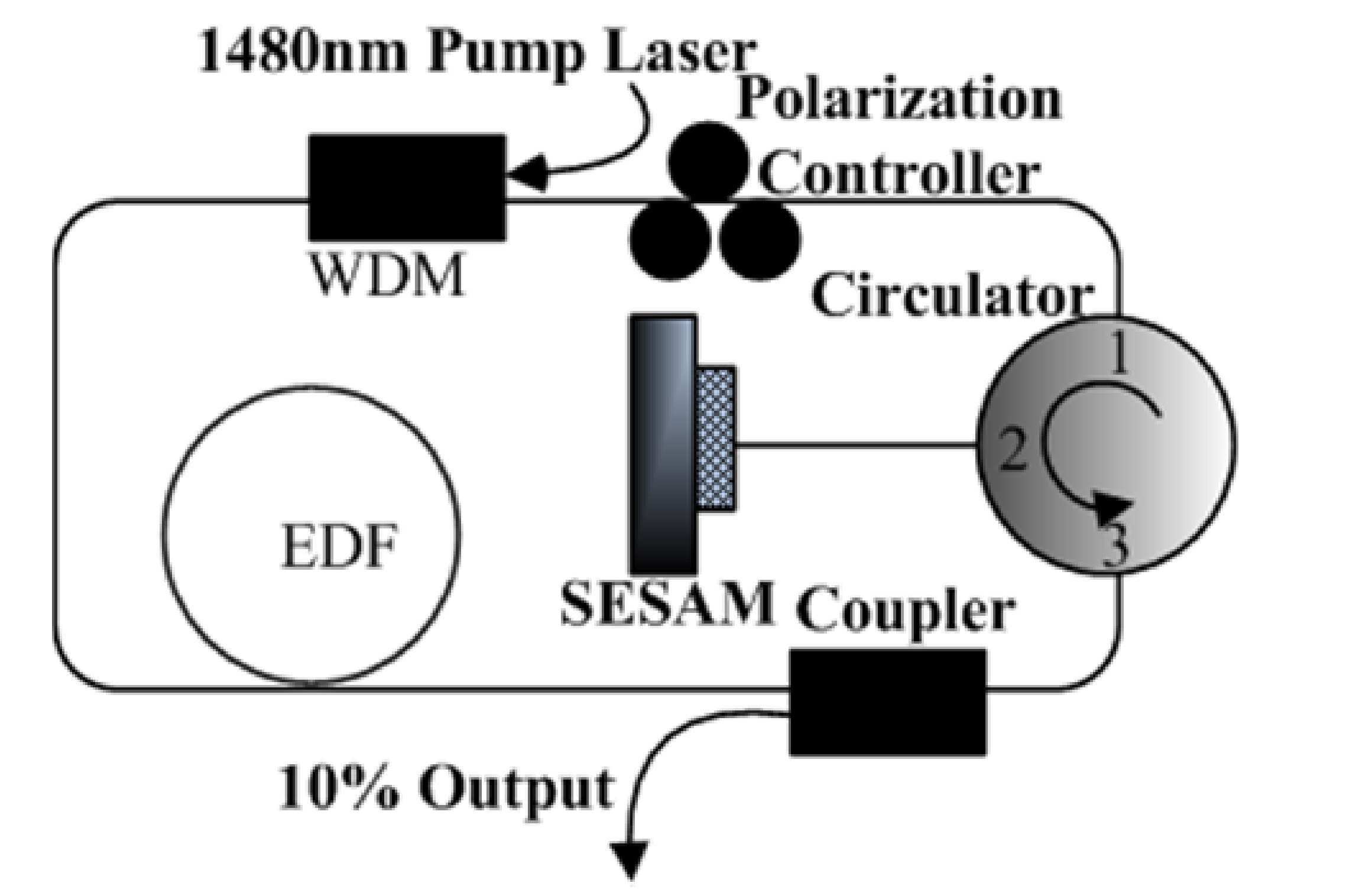

Fig. 1 H. Zhang et al

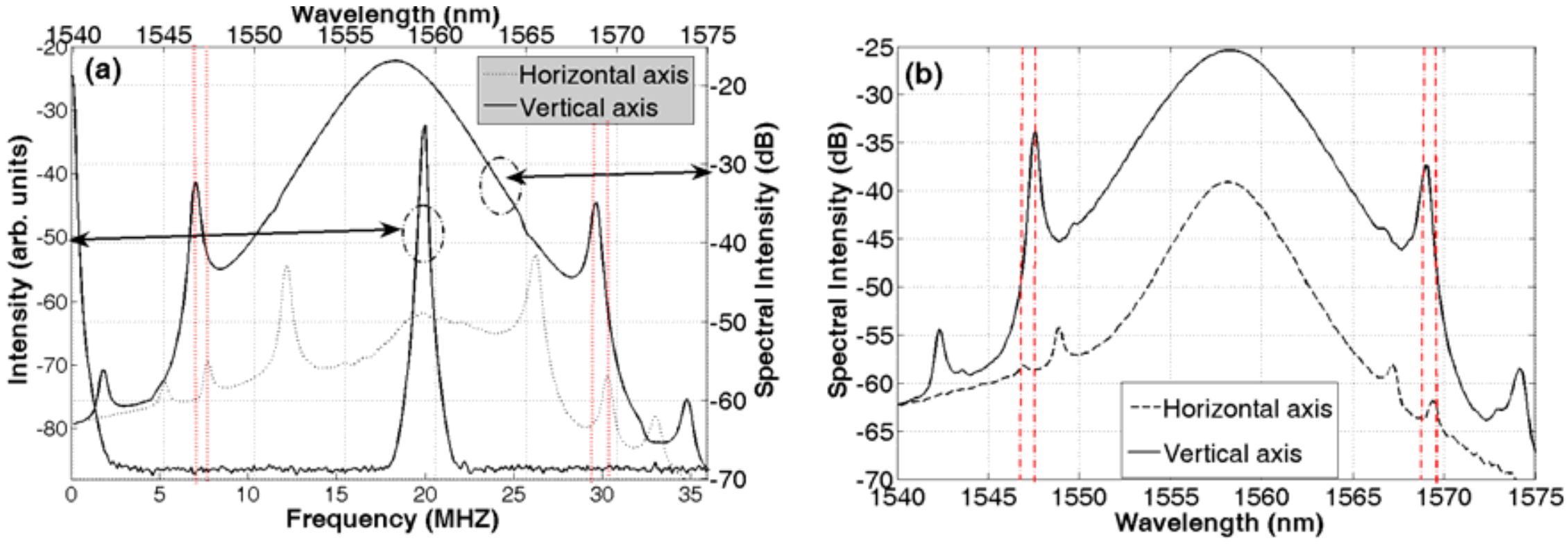

Fig. 2 H. Zhang et al

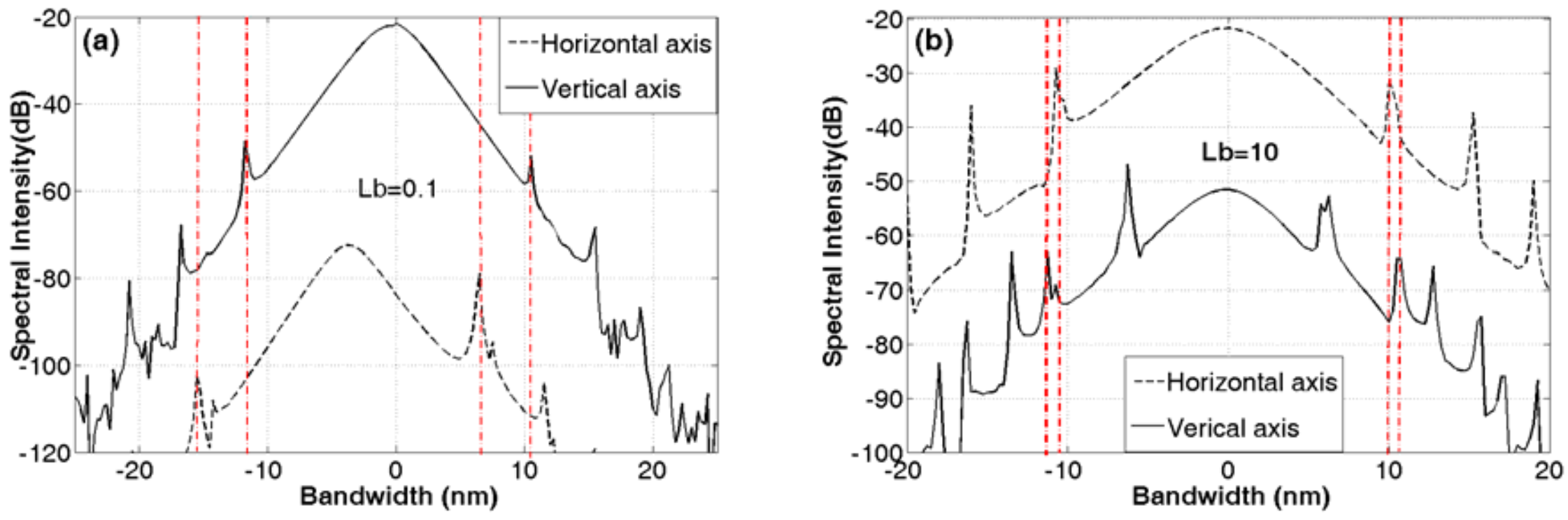

Fig. 3 H. Zhang et al